\documentclass{aa}  

\usepackage{graphicx}
\usepackage{txfonts}
\usepackage{lipsum}
\usepackage{subcaption}         % necessary for continued figures, example in section 3
\usepackage{lscape}             % to rotate a single page table, example in appendix.
\usepackage{placeins}           % useful with \FloatBarrier, to keep 
\newcommand{\cmi}{$\text{cm}^{-1}$}
\newcommand{\kms}{$\text{km s}^{-1}$}
\newcommand{\ia}{$\text{i}^{1}\Pi - \text{a}^{1}\Delta$}
\newcommand{\Toto}{\textsc{Toto}}

\begin{document}

%%%%%%%%%%%%%%%%%%%%%%%%%%%%%%%%%%%%%%%%
% if you use custom commands in your title,
% ensure to check your title when submitting!
%%%%%%%%%%%%%%%%%%%%%%%%%%%%%%%%%%%%%%%%
% \title{Astronomy \& Astrophysics \LaTeX\ template}
% \title{A new singlet system i\,$^1\Pi$ - a\,$^1\Delta$ of the TiO molecule\thanks{Full Tables 1 and 3 are only available at the CDS.}}
%\title{First detection of the TiO  i\,$^1\Pi$ - a\,$^1\Delta$ system in stellar spectra and its laboratory characterization\thanks{Full Tables 1 and 3 are only available at %the CDS.}}
\title{First identification of the TiO  i\,$^1\Pi$ - a\,$^1\Delta$ system in stellar spectra and its spectroscopic characterization\thanks{Full Tables 1 and 3 are only available at the CDS.}}
%\title{Unexpectedly strong singlet system i\,$^1\Pi$ - a\,$^1\Delta$ of the TiO molecule\thanks{Full Tables 1 and 3 are only available at the CDS.}}

%%%%%%%%%%%%%%%%%%%%%%%%%%%%%%%%%%%%%%%%
% \author{M. R. Schmidt\fnmsep\thanks{Corresponding author: schmidt@ncac.torun.pl}}
%\author{M. R. Schmidt\thanks{Corresponding author: schmidt@ncac.torun.pl}}
%\author{M. R. Schmidt\email[show]{schmidt@ncac.torun.pl}}
\author{Mirek R. Schmidt}
\institute{Nicolaus Copernicus Astronomical Center, Polish Academy of Sciences, ul. Rabia{\'n}ska 8, 87-100 Toru{\'n}, Poland\\
\email{schmidt@ncac.torun.pl}}
% orcid 0000-0003-1172-0946

\date{Received --, --}

  \abstract
{
TiO plays an important role in determining the atmospheric structure of M-type stars
and in shaping the visual part of their spectra. 
The precision of synthetic spectra when confronted with high-resolution observations 
depends on the completeness of the included transitions and on the accuracy of line positions and line intensities.
}
   {
% Mandatory. The objectives of the paper are defined here.
We aim to systematically assess the quality of synthetic spectra computed with
up-to-date TiO line lists by comparing them 
with high-resolution, high signal-to-noise spectra of cool stars calibrated in absolute or relative flux.
} 
   {
% Mandatory. The methods of the investigation are outlined here
We compute synthetic spectra for late-type M giants and identify systematic discrepancies 
in the wavelength range 5810--5850\,\AA.
To investigate the origin of these discrepancies, we analyse experimental TiO absorption cross-sections.
}
   {
% Mandatory. The results are summarized here.
We report the detection of a molecular band of the  singlet system  $^{1}\Pi$--a$^{1}\Delta$ of the TiO, 
with an R-head located at 
% $\lambda$
5814.8\,\AA, 
overlapping  the 1--3 band of the $\alpha$  (C\,$^3\Delta$ - X\,$^3\Delta$) system.
The lower state of the band is identified as the a$^1\Delta$ v$^{\prime\prime}$=0 state,
while the upper state is most likely the $^1\Pi$ in its ground vibrational state.
The empirical band intensity is derived by comparing the relative strengths of neighbouring 
bands from the  C\,$^3\Delta$--X\,$^3\Delta$  and B\,$^3\Pi$--X\,$^3\Delta$ 
systems in the experimental cross-section.
The band intensity is further validated 
by synthetic spectrum calculations for the late-type giant
\object{30 Her} (M6 III) and comparison with its observed spectrum from the MELCHIORS library. 
}
   {
% Optional, leave empty if necessary.  “Conclusions” can be used to
% explicit the general conclusions that can be drawn from the paper.
The i\,$^{1}\Pi$ (optionally 2\,$^1\Pi$) electronic state has been predicted by theoretical calculations.
While energies of two vibrationally excited states were previously inferred from de-perturbation analyses of the 
$\alpha$  system, accurate spectroscopic constants for this electronic state were lacking. 
The newly identified band is sufficiently strong to affect the flux distribution in the spectra of cool stars.
}

\keywords{molecular data -- stars: atmospheres -- stars: late-type}

\maketitle

\nolinenumbers

%%%%%%%%%%%%%%%%%%%%%%%%%%%%%%%%%%%%%%%%%%%%%%%%%%%%%%%%%%%%%%
\section{Introduction}

Since its identification in the spectra of  M-type stars
(\citealt{Fowler1904}), TiO has been recognised as a major source of opacity
in the atmospheres of cool stars. The visual spectra of late M-type stars
are shaped by deep absorption features from electronic systems between
low-lying states of TiO, providing key signatures for spectral
classification (\citealt{Merrill1962}).  In high-resolution spectra, TiO
lines strongly affect synthetic spectra, and the analysis of weak atomic
lines or other molecular bands is nearly impossible without prior identifying
numerous TiO transitions (e.g.  the determination of Li I;
\citealt{Woodward2020,Kaminski2023}).  At the same time, molecular bands,
owing to the large number of states spanning a wide range of energies,
provide valuable diagnostics of the dynamical behaviour of different layers
in stellar atmospheres and their surroundings (see e.g. 
\citealt{Tylenda2009}).

To support stellar atmosphere modelling, line lists for TiO and its minor
isotopologues were compiled (\citealt{Jorgensen1994,Plez1998}), initially
based on laboratory measurements.  As quantum-chemical {\it ab initio}
calculations improved (\citealt{Langhoff1997}), theoretically based line
lists became available, including the AMES list (\citealt{Schwenke1998}) and
the more recent \Toto\ list (\citealt{McKemmish2019}), produced within the
ExoMol{\footnote{www.exomol.com} effort to determine molecular opacities in cool astrophysical objects
(\citealt{Tennyson2024}).  

A major advantage of {\it ab initio} methods is that
they provide transition dipole moment functions, spin–orbit couplings, and
other quantities required to compute band intensities, while also enabling
physically motivated extrapolation to high vibrational and rotational levels
(e.g.  up to ($v$=20) and ($J$=400) in \Toto) that contribute significantly to
opacity at elevated temperatures.

Theory also allows estimates of the energies and properties of as yet
unobserved states (see e.g.  \citealt{Dobrodey2001,MM2010}) and of bands
that may be too weak in normal giants but detectable under specific
astrophysical conditions.  For example, the \cite{Schwenke1998} line list
includes the forbidden c$^1\Phi$--X$^3\Delta$ and b$^1\Pi$--X$^3\Delta$ bands,
later identified in the spectrum of the red nova V838 Mon during the
cooling phase after its eruption (\citealt{Kaminski2009}).

Despite these advances, theoretical calculations still do not provide
sufficiently accurate line positions, so {\it ab initio} line lists rely heavily
on precise laboratory data.  Since its astronomical identification, TiO has
been extensively studied in the laboratory and is now among the
best-characterised transition-metal oxides (\citealt{Merer1989}; see e.g. 
\citealt{Phillips1973,Ram1999,Barnes1997}, summarized by
\citealt{McKemmish2017}).  More recent work includes analyses of singlet
transitions by Bernath’s group
(\citealt{HodgesBernath2018,BittnerBernath2018}) and the weak $\epsilon$
band (\citealt{BernathCameron2020}).

A recent study of the TiO $\gamma^\prime$ (B$^3\Pi$--X$^3\Delta$) system by
\cite{BernathBhusalSchmidt2025} analysed the ($\Delta v$=+1) sequence and
derived spectroscopic constants for vibrationally excited levels of the
B$^3\Pi$ state up to ($v$=4).  Using spectra of a late-type giant and a dwarf,
they demonstrated the high accuracy achievable with high-quality data, although
fitting the absolute flux distribution required modifications to the
intensities of excited bands relative to those adopted from the \Toto\ line
list.  For such {\it ad hoc} adjustments to be reliable, they should be
physically justified.  The present work is motivated by a
systematic verification of available molecular line lists in both line
positions and band intensities.

Here we focus on the spectral region blueward of the B$^3\Pi$--X$^3\Delta$
(1--0) system, dominated by TiO lines overlapping with the excited
C$^3\Delta$--X$^3\Delta$ ($\Delta v$=-2) sequence.  We identify and analyse in
detail a band of the previously unknown singlet system \ia.  Section 2
presents the experimental and observational data.  The observational
motivation and analysis of the band, search for other bands of the system
 are given in Sect.  3.  In section  3 we
discuss also the application of the resulting line list to stellar spectra and
the empirical determination of line strengths.  Section 4 provides a broader
discussion and Sect. 5 presents final remarks.

\begin{figure*}[t]
\centering
\includegraphics[width=\textwidth]{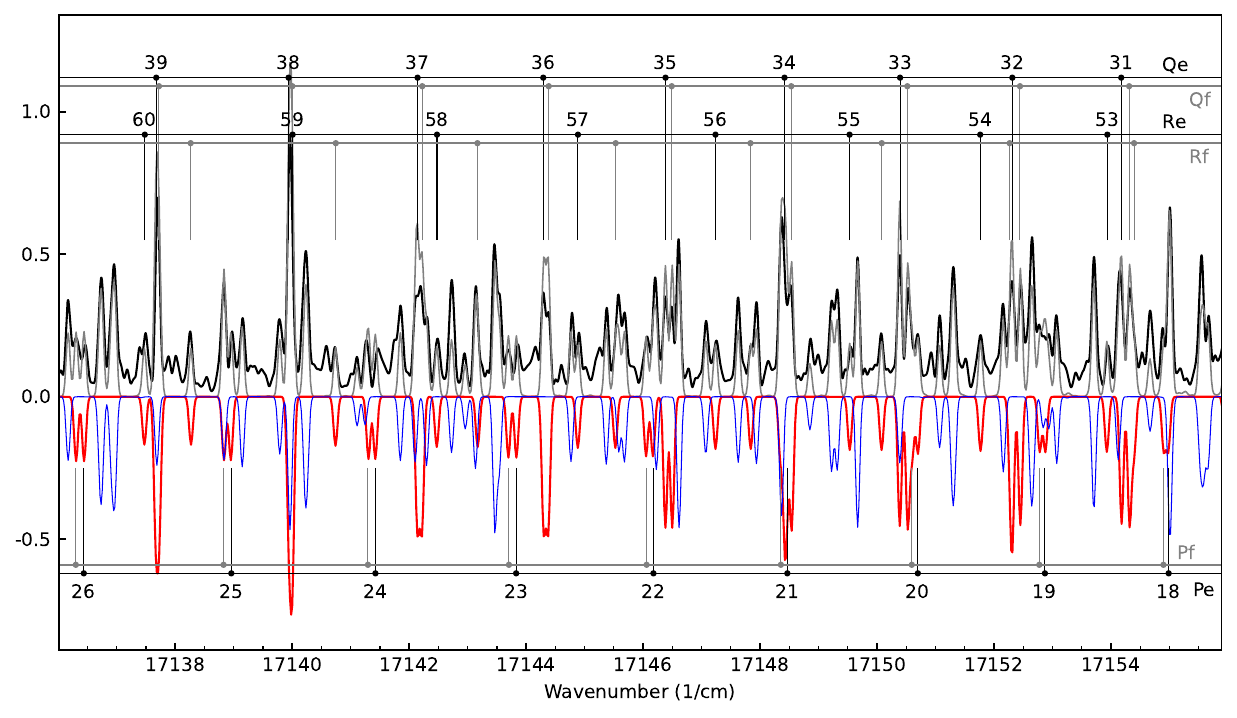}
\caption{
Detailed view of the laboratory spectrum showing features of the \ia\ (0--0) band.
The laboratory spectrum is plotted in black, with the full simulated spectrum overplotted in grey. 
The mirrored spectra illustrate the separate contributions of the C$^3\Delta$--X$^3\Delta$ (0--2) and 
(1--3) bands (blue) and the \ia\ (0--0) band (red). 
Line assignments for individual transitions of the \ia\ band in the P, Q, and R branches, 
for both parities, are indicated above and below the spectrum. 
The rotational quantum numbers J apply to both parities.
}
\label{figure:tio_lab_ia}
\end{figure*}

\section{Spectral data} \label{sec: Data}

The experimental TiO cross-section used in this analysis was described in
detail and made publicly available by \cite{Bernath2020}.  It is based on an
emission spectrum recorded in 1985 at the McMath–Pierce Solar Telescope
using the 1 m Fourier transform spectrometer operated by the National Solar
Observatory at Kitt Peak, Arizona, by S.  Davis, G.  Stark, J.  Wagner, and
R.  Hubbard.  The excitation temperature of the gas was determined to be
($2300 \pm 100$) K by simulating the (0--0) band of the A$^3\Phi$--X$^3\Delta$
transition with PGOPHER (\citealt{Western2017}) using the spectroscopic
constants of \cite{Ram1999}.  The spectral resolution is 0.044\,\cmi,
and the wavenumber scale was calibrated to an accuracy of $\pm$0.002\,\cmi\
(see \citealt{Bernath2020} for details).  This is the same experimental
spectrum used in a series of previous TiO studies
(\citealt{BernathBhusalSchmidt2025,CameronBernath2022,BernathCameron2020}). 
In the following, we refer to this dataset simply as the experimental
spectrum, unless stated otherwise.  A portion of the experimental
cross-section is shown as a solid black line in
Fig.~\ref{figure:tio_lab_ia}.

For the analysis of stellar spectra, we used data from the MELCHIORS library
(\citealt{Royer2024}), a spectral survey of bright northern-hemisphere
stars.  The spectra were obtained with the HERMES spectrograph
(\citealt{Raskin2011}) mounted on the Mercator Telescope at the Roque de los
Muchachos Observatory on La Palma, Spain.  The library provides high
spectral resolution ((R $\approx 85,000$)) and high signal-to-noise ratios,
with spectra corrected for the instrumental response and preserving the
relative spectral energy distribution.  This also applies to the spectrum of
\object{30 Her} (M6 III) used in this work.  Its signal-to-noise ratio in
the visual range is about 97, which is sufficient for our analysis.

\section{Method and results} 
% \label{sec: Results}

 \begin{table*}[h]
 \centering
 \caption{Sample of lines and residuals for the analysed band.}
 \begin{tabular}{cccc rrr c}
 \hline \hline
 $J^{\prime}$ & $e/f^{\prime}$ & $J^{\prime\prime}$ & $e/f^{\prime\prime}$ & Observed & Calculated & Obs - Calc & Line Assignment \\
    &       &     &       & (cm$^{-1}$) & (cm$^{-1}$) &  (cm$^{-1}$) &  \\   
\hline
41 & f  & 41 & e & 17132.9805 & 17132.9768 &  0.0037 & qQe(41)~~   i $^1\Pi$ v=0 41 41 F1f - a$^1\Delta$ v=0 41 41 F1e \\
41 & e & 41 & f  & 17132.9805 & 17132.9816 & -0.0011 & qQf(41)~~   i $^1\Pi$ v=0 41 41 F1e - a$^1\Delta$ v=0 41 41 F1f \\
63 & f  & 62 & f  & 17133.1052 & 17133.1196 & -0.0144 & rRf(62)~~    i$^1\Pi$ v=0 63 63 F1f - a$^1\Delta$ v=0 62 62 F1f \\
26 & e & 27 & e & 17133.8847 & 17133.8544 &  0.0303 & pPe(27)~~   i $^1\Pi$ v=0 26 26 F1e - a$^1\Delta$ v=0 27 27 F1e \\
62 & e & 61 & e & 17134.8949 & 17134.8928 &  0.0021 & rRe(61)~~   i $^1\Pi$ v=0 62 62 F1e - a$^1\Delta$ v=0 61 61 F1e \\
40 & f & 40 & e & 17135.3640 & 17135.3543 &  0.0097 & qQe(40)~~   i $^1\Pi$ v=0 40 40 F1f - a$^1\Delta$ v=0 40 40 F1e \\
40 & e & 40 & f & 17135.3640 & 17135.3809 & -0.0169 & qQf(40)~~   i $^1\Pi$ v=0 40 40 F1e - a$^1\Delta$ v=0 40 40 F1f \\
25 & f & 26 & f & 17136.3032 & 17136.2990 &  0.0042 & pPf(26)~~   i $^1\Pi$ v=0 25 25 F1f - a$^1\Delta$ v=0 26 26 F1f \\
25 & e & 26 & e & 17136.4604 & 17136.4355 &  0.0249 & pPe(26)~~   i $^1\Pi$ v=0 25 25 F1e - a$^1\Delta$ v=0 26 26 F1e \\
61 & e & 60 & e & 17137.4918 & 17137.4788 &  0.0130 & rRe(60)~~   i $^1\Pi$ v=0 61 61 F1e - a$^1\Delta$ v=0 60 60 F1e \\
39 & f & 39 & e & 17137.6973 & 17137.6746 &  0.0227 & qQe(39)~~   i $^1\Pi$ v=0 39 39 F1f - a$^1\Delta$ v=0 39 39 F1e \\
39 & e & 39 & f & 17137.6973 & 17137.7210 & -0.0237 & qQf(39)~~   i $^1\Pi$ v=0 39 39 F1e - a$^1\Delta$ v=0 39 39 F1f \\
61 & f & 60 & f & 17138.2531 & 17138.2628 & -0.0097 & rRf(60)~~   i $^1\Pi$ v=0 61 61 F1f - a$^1\Delta$ v=0 60 60 F1f \\
\hline
\end{tabular}
\label{table:residuals}
\tablefoot{The line assignment illustrate the transition $^{\Delta N}\Delta J(e/f)"_(J'')$ i$^1\Pi$ $v'$ $J'$ $N'$ $F1p'$ - a$^1 \Delta^+$
$v''$ $J''$ $N''$ $F1p"$.
The full table is available at the CDS.}
% \tablefoottext{a}{Parameter held fixed in the fit at value derived by \cite{Kaledin1995}}

% \tablecomments{All values are in cm$^{-1}$. One standard deviation error is indicated in parentheses. 
% The spectroscopic constants for v = 0 and v = 1 levels were obtained from \cite{cameron2022visible}.
% for better visibility we have change designation of states to more publishable than in the original files
% }
\end{table*}

The direct motivation for searching for a new band was the observation of
% a deficit in 
extra
opacity in the spectra of late-type giants within a spectral
region that almost exactly coincides with the $\alpha$
(C$^3\Delta$--X$^3\Delta$) (1--3) band, which extends redward from its band
head at 5809.9\,\AA.  This discrepancy becomes evident when synthetic spectra
are compared with observations.  An example is shown in
Fig.~\ref{fig:30her_cx_melchiors} for the MELCHIORS spectrum of the
late-type giant \object{30 Her} (M6 III).  

The synthetic spectrum calculated with an available list of lines
is above the flux in the observed spectrum.
Details of the fitting procedure are discussed in text below.

The mismatch cannot be explained by adjusting the relative intensities of
the C$^3\Delta$--X$^3\Delta$
(0--2) and (1--3) bands.  Even with
modified band strengths, the synthetic spectrum fails to reproduce the fine
structure of
the observations.  Similar results were obtained for other
late-type giants.  In contrast, when the stellar spectrum was overplotted
with the experimental TiO spectrum, nearly all features in this wavelength
region showed close correspondence.  This strongly suggested that the
missing opacity originates from a TiO band that is poorly represented or
absent in existing line lists.

The stellar spectrum alone provides only the position of the apparent band
head of the unknown system, measured at (5814.8$\pm$0.1)\,\AA\ 
or
(17,192.73$\pm$0.3)\,\cmi.\footnote{Wavelengths are given in air, and wavenumbers in
vacuum.}

\subsection{Analysis of the laboratory spectrum}

The laboratory spectrum was analysed using the molecular spectroscopy 
fitting program PGOPHER \citep{Western2017}.  The overlapping bands 
of the $\alpha$ system 
were simulated using spectroscopic constants 
from \cite{HodgesBernath2018}. 
To reproduce line positions accurately at high $J$, small adjustments to the
spectroscopic constants of the C$^3\Delta$ $v$=0 and $v$=1 states were
required; only the higher-order correction terms ($A_\mathrm{H}$) were
modified.

A detailed inspection of the laboratory spectrum revealed several relatively
strong unassigned line pairs exhibiting characteristic $\Lambda$-splitting,
later identified as Q(e/f)-branch transitions.  Subsequent analysis
uncovered sequences belonging to the P and R branches of both parities.  The
Q-branch lines are approximately twice as strong as those of the P and R
branches, and the observed line splittings indicated that the system
involves a $\Pi$ state.  The stellar position of the band head provided a
constraint on the location of the R head, which is not clearly distinguished
in the laboratory spectrum.

After several iterations of assigning rotational quantum numbers $J$ to
individual lines, and allowing spectroscopic parameters of both the upper
and lower states to vary simultaneously, a satisfactory fit was obtained. 
In total, 161 transitions assigned to the investigated band were fitted
using the \textbf{N$^2$} Hamiltonian, covering $J$ from 7 to 54 in the P branch,
$J$ from 21 to 69 in the R branch, and $J$ from 10 to 77 in the Q branch.  The 
% unweighted
standard deviation of the fit is 0.018\,cm$^{-1}$.  The resulting
spectroscopic parameters are listed in Table~\ref{table:spec_constants}. 
The absolute energy of the a$^1\Delta$ state was fixed at the value derived
from analysis of the forbidden C$^3\Delta$--a$^1\Delta$ system
\citep{Kaledin1995}.
A detailed list of fitted transitions is given in
Table~\ref{table:residuals}.

Allowing the lower-state parameters to vary led to two possible solutions. 
The preferred solution identifies the lower state as a$^1\Delta$, as
presented in Table~\ref{table:spec_constants}.  An alternative solution
involved the X$^3\Delta$, $\Omega=3$ spin component \citep{Ram1999}.  This
possibility was excluded based on the astrophysical arguments discussed in
Sect. 4.

The spectroscopic parameters of the upper state do not match those of any
previously well-characterised electronic state, either in band origin or
rotational constant.  We therefore conclude that the observed band belongs
to a previously unrecognised singlet system, $^1\Pi$--a$^1\Delta$.  Following
the convention adopted in the \Toto/ExoMol TiO line list
\citep{McKemmish2019}, we designate the upper state as i$^1\Pi$.
We further assume that the upper state is the $v$=0 vibrational level. 
The justification for this assumption is given in Sect. 4.

%=============================================================================================================
 \begin{table*}[ht]
        \centering
        \caption{Spectroscopic constants for the i$^1\Pi$ and a$^{1}\Delta$ $v$=0 states of TiO.}
        \begin{tabular}{lllll}
            \hline \hline
                                                    &   i $^1\Pi$ $v$=0     &  a $^1\Delta$ $v$=0  & &  \\
                                                    &                              &  (this work)             & Bittner \& Bernath (2018)  & Amiot et al. (1996) \\
             \hline
            T$_v$                              &   20627.5192(38) &  3444.367\tablefootmark{a}     &                                & \\
            B$_v$                              &   0.507108(60)     &  0.536255(61)        &  0.53624855(795)   & 0.5362187(24)   \\
            D$_v \times 10^{7}$        &   6.07(11)             &  6.13(11)                 &  6.0293(115)           & 5.9914(24)        \\
            H$_v \times 10^{10}$      &                             &                                &                                &                           \\
            q$\times 10^4$                &   3.683(85)           &                                &  & \\
            q$_D \times 10^{7}$        &  -2.625(58)           &                                &  & \\
            q$_H \times 10^{11}$      &   2.916(91)           &                                &  & \\
            \hline
        \end{tabular}
        \label{table:spec_constants}
        \tablefoot{All values are in cm$^{-1}$. One standard deviation error is indicated in parentheses. 
        \tablefoottext{a}{Parameter held fixed in the fit at value derived by \cite{Kaledin1995}}
        }
        %\tablefootmark{a}{fixed value}
\end{table*}

In fact, the upper i$^1\Pi$ state is not entirely unknown.  Vibrationally
excited levels of this state were reported by \cite{Namiki2003} in their
study of perturbations affecting excited C$^3\Delta$ levels observed in
$\alpha$-system bands, based on laboratory data from \cite{Phillips1973}. 
They derived spectroscopic parameters for two consecutive vibrational levels
of the perturbing state, although their absolute vibrational numbering
remained uncertain.

With the adopted
identification of the observed level as the $v$=0 state of
i$^1\Pi$, we assign \cite{Namiki2003} levels to $v$=2 and $v$=3.
As noted
by \cite{Namiki2003}, the $v$=3 level exhibits irregular behaviour, with an
increasing rotational constant $B_v$, which they attributed to interaction
with the as yet unobserved h$^1\Sigma^{+}$ state.  Their value for the $v$=2
rotational constant, $B_v = 0.5018 \pm 0.0044$ cm$^{-1}$ (their Table 5),
can be directly compared with the value derived here.
Assuming $\alpha_e = 3\times10^{-3}$ cm$^{-1}$, typical for TiO electronic
states to within about 10\% (see Table 5 of \citealt{MM2010}), and using relation
$B_v = B_e - \alpha_e (v + 1/2)$ to derive $B_e = 0.508608$ cm$^{-1}$ from$B_v$ for $v$=0, we obtain
$B_v = 0.501108$ cm$^{-1}$ for $v$=2, in agreement within uncertainties with
the value of \cite{Namiki2003}.

Figure~\ref{figure:tio_lab_ia} shows a portion of the laboratory spectrum
% with the identified \ia\ transitions marked, 
together with PGOPHER
simulations of the \ia\ (0--0) band and the overlapping C$^3\Delta$--X$^3\Delta$ bands.  The displayed
region contains the strongest lines of the \ia\ system.

The sign of the $q$ parameter was determined using the pure precession and
unique perturber approximations, in which the $\Lambda$-doubling parameter
of a state of energy $E$ perturbed by a state at energy $E_{\mathrm{pert}}$
is approximated by
\begin{equation}
      q = \frac{4 B^2}{E - E_{\mathrm{pert}}}
\end{equation}
where $B$ is the rotational constant of the perturbed state
\citep{LeFebvreBrionField1986}.  Applying this relation to the b$^1\Pi$ and
i$^1\Pi$ states allows an estimate of the energy of the perturbing state. 
Using $T(v=0)$, $q$, and $B$ for b$^1\Pi$ (14\,716 cm$^{-1}$,
$-1.6325\times10^{-4}$, and 0.51204209 cm$^{-1}$, respectively
\citep{BittnerBernath2018} and the corresponding parameters for i$^1\Pi$
from Table~\ref{table:spec_constants}, we derive an estimated perturber
energy of 18\,812 cm$^{-1}$.

This is consistent with theoretical predictions for the unobserved singlet
state h$^1\Sigma^{+}$: \cite{MM2010} obtained $T_\mathrm{e} = 18\,596$
cm$^{-1}$ at the MRCI+Q level (their state 2\,$^1\Sigma^{+}$), while
\cite{Schwenke1998} reported $T_\mathrm{e} = 17\,564.88$ cm$^{-1}$.  The
perturbed states and the perturber differ mainly by $\pi\delta$ versus
$\delta^2$ configurations (\citealt{Schwenke1998,Merer1987}), satisfying the
assumptions of the approximation.  This analysis supports the adopted
assignment of e and f parities.

The laboratory position of the R head of the i$^1\Pi$--a$^1\Delta$ (0--0) band
corresponds to the interpolated position of the R$_e$(17) line at
17\,192.524 cm$^{-1}$ (5814.88 Å), while the Q branch begins with Q$_f$(2)
at 17\,182.979 cm$^{-1}$ (5818.10 Å).

\subsection{Search for other bands}

Encouraged by the successful analysis of the (0--0) band 
we searched for the next band in the $\Delta v$=0 sequence.

To this end, we estimated the vibrational constants $\omega_{\mathrm{e}}$
(and $\omega_{\mathrm{e}}x_{\mathrm{e}}$) using the band origins derived in
this work together with those reported by \cite{Namiki2003}.  A linear fit
to the origins of the $v$=0, 2, and 3 levels gives $\omega_{\mathrm{e}} =
841$\,\cmi.  Including the anharmonic term in the expansion does not change
the result, as the derived $\omega_{\mathrm{e}}x_{\mathrm{e}}$ is close to
zero.  Fixing the anharmonic term at a value typical for other electronic
states, $\omega_{\mathrm{e}}x_{\mathrm{e}} = 4$\,\cmi, yields
$\omega_{\mathrm{e}} = 857$\,\cmi.

Theoretical calculations by \cite{MM2010} give $\omega_{\mathrm{e}} = 924$\,\cmi\ 
and $\omega_{\mathrm{e}}x_{\mathrm{e}} = 4$\,\cmi.
 
One may also estimate $\omega_{\mathrm{e}}$ using the approximate formula 
(exact for a Morse potential)
\begin{equation}
D_v = \frac{4 B^3}{\omega_{\mathrm{e}}^2}.
\end{equation}
This yields $\omega_{\mathrm{e}} = 933$\,\cmi\ with an uncertainty of about
15\,\cmi, arising from the uncertainty in $D_v$.  Overall, the value of
$\omega_{\mathrm{e}}$ remains poorly constrained.

These estimates allowed us to narrow the search range for the \ia\ bands (1--1)
and (1--0) in the experimental spectrum.  In particular, we searched for
characteristic patterns formed by Q-branch lines ($J$=38--41), where the
expected $\Lambda$-doubling is close to zero and lines of both parities
overlap, forming features of approximately twice the usual intensity.  So
far, however, we have been unable to reliably identify lines belonging to
the (1--1) or (1--0) bands.
Therefore the value of $\omega_{\mathrm{e}}$ remains uncertain.

The identification of lines belonging to the (0--1) and (0--2) bands in the
laboratory spectrum was also unsuccessful, despite their positions being
well constrained using spectroscopic parameters of excited levels of
a$^1\Delta$ from \cite{BittnerBernath2018}.  The R-head positions of the (0--1)
and (0--2) bands are expected at 6177.116\,\AA\ ($^{\mathrm{R}}$R$_e$(19),
16\,184.307\,\cmi) and 6583.435\,\AA\ ($^{\mathrm{R}}$R$_e$(21),
15\,185.329\,\cmi), respectively.  The (0--1) band lies in the crowded region
of the B$^3\Pi$--X$^3\Delta$ (0--0) band.

Because the experimental spectrum represents an absorption cross-section at
an excitation temperature of 2300~K, the populations of the a$^1\Delta$
% $v^{\prime\prime}=1$ 
$v$=1 and 
% $v^{\prime\prime}=2$
$v$=2 levels are reduced by factors
of approximately 2 and 4, respectively, relative to $v$=0,
further decreasing the chances of their identification.

\subsection{Empirical intensities}

\begin{figure*}
\centering
\includegraphics[width=\textwidth]{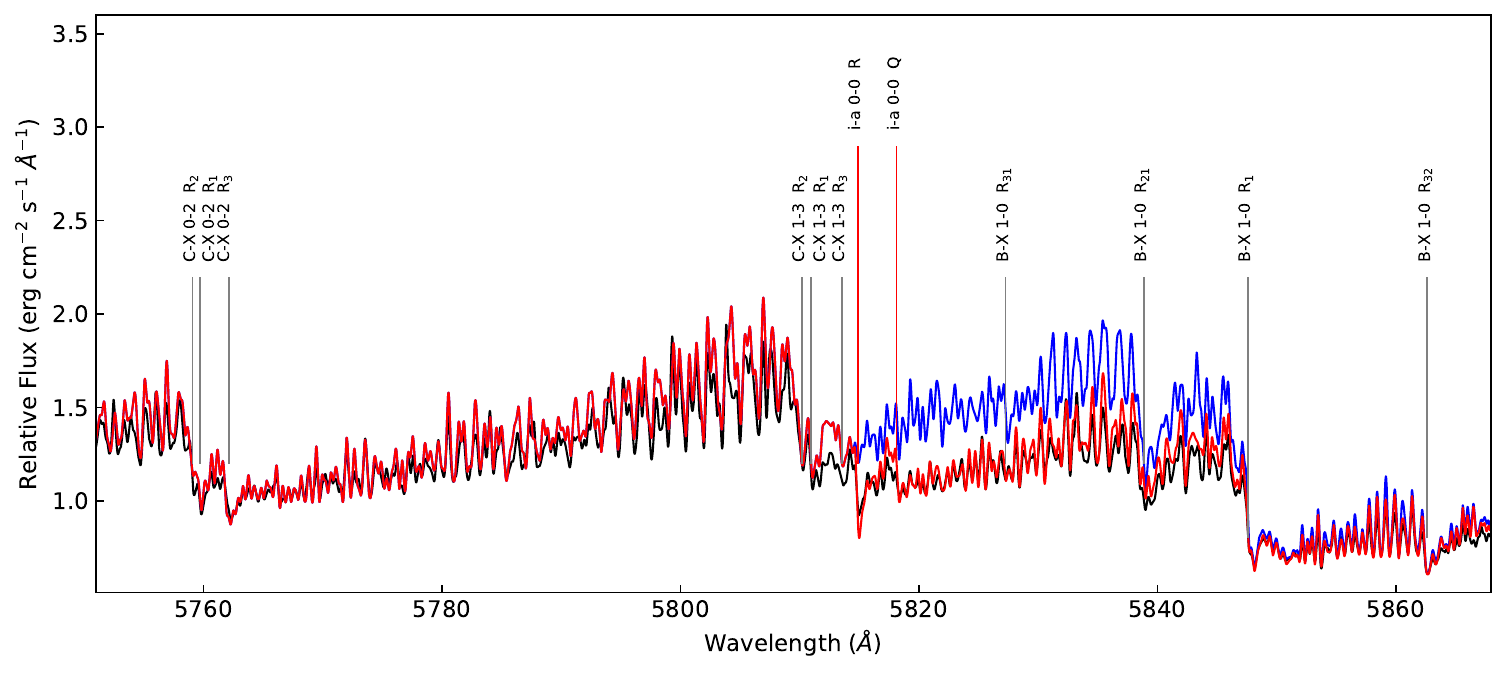}
\caption{MELCHIORS spectrum of  \object{30 Her} (black) with over-plotted synthetic spectra
calculated for updated \Toto\ list of lines described in text without (blue) and with (red) lines of the \ia\ (0--0) band
%  (blue) and with lines of the 
% The atmospheric model has lower T$_{eff}$ 3000 K, log\,g=0.0, Z=0 and v$_{t}$=4\,\kms.
}
\label{fig:30her_cx_melchiors}
\end{figure*}

The empirical intensities of lines in the \ia\ 0–0 band were determined in
two ways.  First, we estimated the band strength from simulations of the
experimental cross-section performed with PGOPHER.  The spectroscopic
constants for the C$^3\Delta$–X$^3\Delta$ (0--2) and (1--3) bands were taken from
\cite{HodgesBernath2018}.  The B–X bands were simulated using data from
\cite{CameronBernath2022} and \cite{BernathBhusalSchmidt2025}.  The strength
parameter for each simulated band in PGOPHER was fixed to reproduce
oscillator strengths from the \Toto\ line list \citep{McKemmish2019}.  With
this approach, the relative strengths of rotational lines within each band
follow the H\"onl–London factors calculated by PGOPHER from effective
Hamiltonians.

The strength parameter of the \ia\ band was then adjusted to match the
laboratory spectrum, and oscillator strengths of individual lines were
derived.  The resulting spectrum, together with individual contributions of
overlapping bands, is shown in Fig.~\ref{figure:tio_lab_ia}.  In the
simulations, the excitation temperature was fixed at 2300\,K, and individual
lines were broadened with a Gaussian profile of width 0.078\,\cmi.

In the next step, we calculated a synthetic spectrum of a late-type giant. 
For this purpose, we used a TiO line list (including isotopologues) based on
the \Toto\ line list \citep{McKemmish2017,McKemmish2019}.  The list was
modified by replacing lines of selected bands of the $\gamma^\prime$ system
with those from the recent analysis of \cite{BernathBhusalSchmidt2025}. 
Only bands with $v^{\prime}, v^{\prime\prime} \leq 6$ were
replaced.\footnote{Because vibrational and spin quantum numbers show some
non-monotonic behaviour for $J>100$, it was first necessary to modify state
assignments for the upper and lower levels in the original states file; see
\cite{Yurchenko2016} for discussion of this issue.}

This list was extended with the \ia\ line list for the main isotopologue and
corresponding lines of minor Ti isotopes.  Isotopic corrections were adopted
from \cite{McKemmish2017}, as implemented in the \Toto\ list.  The
isotopic shifts were derived from their i$^1\Pi$ and a$^1\Delta$ states. 
Although the i$^1\Pi$ energies calculated by \cite{McKemmish2017} are higher
than the observed values reported here, their effect on the isotopic
corrections was neglected.

As a model atmosphere for 30\,Her, we adopted a MARCS model with standard
chemical composition, solar metallicity, T$_{\mathrm{eff}} = 3000$\,K,
log\,$g = 0.5$, and $v_{\mathrm{t}} = 4$\,\kms\ (\citealt{MARCS}).  The
adopted temperature is somewhat lower than expected for an M6\,III star. 
The primary motivation for this choice was to achieve the best simultaneous
fit to neighbouring bands.  Using higher T$_{\mathrm{eff}}$ values makes it
difficult to reproduce both the $\alpha$ and $\gamma^\prime$ systems
simultaneously with the adopted line list.

We find that no additional adjustment of the \ia\ band strength is required
to reproduce the spectrum of the giant.  The results of the analysis are
illustrated by the synthetic spectrum overplotted on the MELCHIORS spectrum
of \object{30\,Her} in Fig.~\ref{fig:30her_cx_melchiors}.  Inclusion of the
i–a band fills the previously missing opacity without requiring any
modification of the intensity of the C$^3\Delta$--X$^3\Delta$ (1--3) band.

An underestimation of opacity in a narrow range around 5812\,\AA\ remains unexplained. 
This position coincides with one of the heads of the ScO A$^2\Pi$--X$^2\Sigma^{+}$ (1--0) band, but
its predicted intensity is insufficient to account for the discrepancy. 
Some mismatches in the region blueward of the B$^3\Pi$--X$^3\Delta$ (1--0)
R$_{21}$ head of the satellite band may be due to the unanalysed (1--1) band of
the \ia\ system.  The contribution of the VO C$^4\Sigma$--X$^4\Sigma$ (1--1)
band near 5786\,\AA\ was neglected in this figure.

The accuracy of the derived line strengths is estimated to be uncertain by
up to 15\%.  Because the potential energy curve of the upper state is not
known, the Franck–Condon factors cannot be determined independently.  The
oscillator strength of the (0--0) band derived from the fit is $f_{00} =
0.082$.  The non-detection of the (0--1) and (0--2) bands provides upper limits on
their strengths of $f_{01} \leq 0.09$ and $f_{02} \leq 0.09$.

The empirical line list for the \ia\ (0--0) band of the main isotopologue
$^{48}$Ti$^{16}$O, based on the derived spectroscopic constants and
empirical intensities extrapolated up to $J$=200, is available as
supplementary material.  A sample of the list with abbreviated line
assignments is shown in Table~\ref{table:linelist}.

% \begin{longrotatetable}
\begin{table*}[ht]
% \begin{sidewaystable*}
\centering
% \begin{deluxetable}{ccccc ccccc}

\caption{Sample table for the PGOPHER line list for the i$^1\Pi$ - a$^1\Delta$ transition of  TiO.}
\begin{tabular}{cccccccccc}
\hline\hline
$J^\prime$ & $P^\prime$ & $J^{\prime\prime}$ & $P^{\prime\prime}$ & Position & $E_{upp}$ &  $E_{low}$ & A &f& Line Assignment \\
 & & & & (cm$^{-1}$) & (cm$^{-1}$) & (cm$^{-1}$) & (s$^{-1}$) & & \\
\hline
%199 & f & 200 & f & 15118.894 & 39132.048 & 24013.154  & 2,800,534.3 & 0.0183 & pPf(200) : i1Pi v=0 199 199 F1f - a1delta v=0 200 200 F1f \\
%198 & f & 199 & f & 15154.328 & 38972.596 & 23818.268  & 2,820,517.5 & 0.0183 & pPf(199) : i1Pi v=0 198 198 F1f - a1delta v=0 199 199 F1f \\
%199 & f & 200 & f & 15118.894 & 39132.048 & 24013.154  & 2,800,534.3 & 0.0183 & pPf(200) \\
%198 & f & 199 & f & 15154.328 & 38972.596 & 23818.268  & 2,820,517.5 & 0.0183 & pPf(199) \\
%
% 63 & e & 62 & e & 17132.249 & 22661.877 & 5529.628 & 3785791.4 & 0.0196 & rRe(62) : i1Pi v=0 63 63 F1e - ... \\
 %41 & f & 41 & e & 17132.977 & 21498.958 & 4365.981 & 8000738.8 & 0.0409 & qQe(41) : i1Pi v=0 41 41 F1f - ... \\
%
 63 & e & 62 & e & 17132.249 & 22661.877 & 5529.628 & 3785791.4 & 0.0196 & rRe(62) : i1Pi v=0 63 63 F1e - a1Delta .. \\ % v=0 .. \\ % 62 62 F1e \\
 41 & f & 41 & e & 17132.977 & 21498.958 & 4365.981 & 8000738.8 & 0.0409 & qQe(41) : i1Pi v=0 41 41 F1f - a1Delta .. \\ % v=0 .. \\ % 41 41 F1e \\
 41 & e & 41 & f & 17132.982 & 21498.963 & 4365.981 & 8000745.5 & 0.0409 & qQf(41) : i1Pi v=0 41 41 F1e - a1Delta .. \\ % v=0 41 41 F1f \\
 63 & f & 62 & f & 17133.120 & 22662.747 & 5529.628 & 3786368.7 & 0.0196 & rRf(62) : i1Pi v=0 63 63 F1f - a1Delta .. \\ % v=0 62 62 F1f \\
 26 & f & 27 & f & 17133.715 & 20983.141 & 3849.426 & 4545768.7 & 0.0224 & pPf(27) : i1Pi v=0 26 26 F1f - a1Delta .. \\ % v=0 27 27 F1f \\
 26 & e & 27 & e & 17133.854 & 20983.280 & 3849.426 & 4545879.5 & 0.0224 & pPe(27) : i1Pi v=0 26 26 F1e - a1Delta .. \\ % v=0 27 27 F1e \\
\hline
\end{tabular}
\tablefoot{J is the total angular momentum, P is the rotationless parity, Position is the calculated line
position in $\mathrm{cm^{-1}}$, $\mathrm{E_{up}}$ and $\mathrm{E_{low}}$ are
the upper and lower energy levels in $\mathrm{cm^{-1}}$, $A$ is the Einstein
$A_{J^\prime \leftarrow J^{\prime \prime}}$ value in s$^{-1}$, and
$f_{J^\prime \leftarrow J^{\prime \prime}}$ is the oscillator strength, and
line assignments are the associated quantum numbers for the given
transition. 
The line assignment illustrate the transition $^{\Delta N}\Delta J(e/f)"(J'')$ : i$^1\Pi$ $v'$ $J'$ $N'$ $F1p'$ - a$^1 \Delta$
$v''$ $J''$ $N''$ $F1p"$.
%In machine readable form the line assignment illustrate the transition $^{\Delta N}\Delta J_{
%(F'_i)F_j''}(J'')$ i$^1\Pi$ v$'$ $J'$ $N'$ $F1p$ - a$^1 \Delta^+$
%v$''$ $J''$ $N''$ $F1p$.
The full table is available at the CDS.}
\label{table:linelist}
% \end{longrotatetable}
\end{table*}
%\end{sidewaystable*}

\section{Discussion}

The next $^1\Pi$ electronic state of TiO above b$^1\Pi$ was predicted
several decades ago (see the summary by \citealt{Merer1989}). 
Quantum-chemical {\it ab initio} calculations have since provided estimates of
its energy (\citealt{Dobrodey2001,MM2010}).  The value closest to that
derived in this work is that of \cite{MM2010}, who obtained T$_\mathrm{e} =
21,100$\,\cmi\ for the i$^1\Pi$ (their 2$^1\Pi$) state at the MRCI+Q level of theory.
Based on these calculations, we infer that the observed upper state is in its ground vibrational mode.

The transition dipole moment of the \ia\ system was estimated by
\cite{Dobrodey2001}, who obtained a value of 0.9 a.u.  near the equilibrium
internuclear distance (3.1 a.u.) (see their Figure~8).  However, the predicted
energy for the i$^1\Pi$ state was 27 052\,\cmi\, significantly higher
than the 20 627\,\cmi\ determined here.  Using their theoretical
transition energy ($\nu_{v v'} = 23 600$)\,\cmi), this corresponds to an
oscillator strength of f$_{00} = 0.036$ (see Eq.  1 of
\citealt{Dobrodey2001}).

The identification of the \ia\ band increases the number of rotationally
analysed singlet states of TiO to seven.  The i$^1\Pi$ state was the last
low-lying singlet state of the $\delta\pi$ configuration with previously
undetermined energy (\citealt{Merer1987,Merer1989}), alongside A$^3\Phi$,
B$^3\Pi$, and c$^1\Phi$.

When the presence of the unidentified band of TiO was first confirmed in stellar
spectra, plausible candidates included forbidden bands of the
D$^3\Sigma^{-}$--X$^3\Delta$ (5--0 or 6--0) and h$^1\Sigma$--X$^3\Delta$ (0--0)
systems.  Fully coupled 
% {\it ab initio} 
calculations by \cite{Schwenke1998}
predicted that these bands could contribute in the analysed spectral
region.\footnote{The NASA AMES TiO line list was distributed for stellar
atmosphere calculations on Kurucz CD-ROM No.  24 \citep{KuruczTiO}.} None of
these systems, however, has yet been observed experimentally.  Because 
{\it ab initio} calculations provide only approximate band positions, some
flexibility existed in assigning the observed lines to these transitions. 
Indeed, the fitted lower-state spectroscopic constants were also consistent
with the X$^3\Delta_3$ state (\citealt{Ram1999}).  A counterargument was
that, if the band belonged to the D$^3\Sigma^{-}$--X$^3\Delta$ or
h$^1\Sigma$--X$^3\Delta$ systems, a stronger contribution from the
X$^3\Delta_2$ spin component would be expected in the lower state.

The hypothesis that the lower state of the band is the ground X$^3\Delta$
state can also be tested using astrophysical observations alone.  Because
the proposed transition would originate from the ground state and overlaps
with the C$^3\Delta$--X$^3\Delta$ band arising from the excited vibrational
level ($v$=3), their relative contributions to opacity should differ in
environments with enhanced columns of cold absorbing gas.  Such conditions
were present in the red nova V838 Mon shortly after its 2002 eruption, when
large column densities of molecular gas at temperatures of 200–300 K were
observed along the line of sight (\citealt{Kaminski2009}).  Although a
strong velocity gradient in the outflow smeared out detailed band structure,
we estimated the column density of gas in the ($v$=3) level of the ground
state by fitting the $\gamma$ (2–-3}) band in the near-infrared.  The same
column density was then used to synthesize the C$^3\Delta$--X$^3\Delta$ (1--3)
band.  This purely observational test showed that the band contour is
insensitive to the presence of large columns of cold gas, effectively ruling
out the ground X$^3\Delta$ state as the lower level of the analysed band.

With the firm identification of a$^1\Delta$  ($v$=0), as the lower state,
these earlier considerations are no longer required.  Nevertheless, the
question of the non-detection of the predicted D$^3\Sigma^{-}$--X$^3\Delta$
and h$^1\Sigma$--X$^3\Delta$ systems remains open.

\section{Conclusions}

The identification of the \ia\ band in stellar spectra has long been
obscured because it overlaps with the much stronger C$^3\Delta$--X$^3\Delta$
(1--3) band and because the lower levels of the two transitions
have similar energies (a$^1\Delta$, ($v$=0): 3444\,\cmi; X$^3\Delta$,
($v$=3): 2973\,\cmi; \citealt{Ram1999}).  As a result, over the wide
temperature range encountered in stars of different spectral types, the
populations of these lower levels vary in a similar way, preserving the
overall contour and masking the presence of the weaker system.

Line intensities of TiO electronic systems may differ by up to an order of
magnitude, as can be seen from comparisons of the $\gamma^\prime$
(B$^3\Pi$--X$^3\Delta$) system in the line lists of \cite{Schwenke1998},
\cite{Plez1998}, and \cite{McKemmish2017}.  Even smaller discrepancies in
other systems can produce noticeable differences in synthetic spectra
(\citealt{Jones2023}), and substantial variations are expected between grids
of late-type stellar models computed with different TiO line lists
(\citealt{Allard2000}).  Observed mismatches in TiO band strengths often
encourage empirical adjustments to band intensities in order to reproduce
stellar spectra.  Within a single electronic system, relative band
intensities are partly constrained by laboratory analyses, for example
through reconstruction of potential energy curves from vibrational data. 
Residual discrepancies may be attributed to inaccuracies in {\it ab initio}
transition dipole moment functions, which is not unexpected.  In the present
case, we encountered a similar issue for the ($\Delta v$=-2) sequence of the
$\alpha$ system; however, the solution proved different.  The discrepancies
were instead caused by the presence of a previously unassigned electronic
system.  This discovery demonstrates that empirical determinations of band
strengths remain uncertain until all significant opacity sources are
identified.

Several TiO systems in the visual spectral range are still poorly
characterized, both in line positions and intensities.  Examples include the
triplet D$^3\Sigma^{-}$--X$^3\Delta$ and the forbidden
h$^1\Sigma$--X$^3\Delta$ systems.  Forbidden bands of c$^1\Phi$--X$^3\Delta$
and b$^1\Pi$--X$^3\Delta$ were identified by \cite{Kaminski2009} in the
spectrum of the red nova V838 Mon, where enhanced column densities of cool
gas made their detection possible.  Notably, the band heads of the
c$^1\Phi$--X$^3\Delta$ system had already been reported as unidentified
features in Mira spectra by \cite{Merrill1962}.  Although laboratory
measurements provide accurate line positions for these bands, their
intensities are currently based only on theoretical calculations
(\citealt{Schwenke1998}) and require experimental verification
(\citealt{Kaminski2009}).

Weak, previously unidentified TiO systems likely contribute only modestly to
the total opacity but can provide additional absorption in spaces between rotational
lines of stronger overlapping bands lowering effectively the level of continuum,
%This reduces the apparent
%quasi-periodic modulation of flux with wavelength, 
an effect that is particularly evident in spectral synthesis of high-resolution 
% spectra where individual lines are narrow.  
In this respect, such systems play a role similar to that of minor
TiO isotopologues (\citealt{Jorgensen1994}).

A systematic approach combining high-resolution, high signal-to-noise
stellar spectra with detailed laboratory analyses can therefore aid in
identifying yet unobserved TiO systems and in reducing discrepancies between
observed and synthetic spectra.  This is an essential step toward more
reliable determinations of stellar parameters and chemical abundances in
M-type stars.

\begin{acknowledgements}
I gratefully acknowledge the memory of my late colleague Dr Yakiv Pavlenko, 
whose collaboration inspired me to continue research on molecular bands in stellar spectra.
I also thank Dr. Tomasz Kamiński for his financial support of M.S.’s participation in the ExoMol 
workshop funded by Grant SONATA BIS No. 2018/30/E/ST9/00398 from 
the Polish National Science Center, and for his patient attention to the early, 
albeit erroneous, hypotheses on the origin of the molecular band analysed here.
\end{acknowledgements}

% Please note that we have included the references below in
% order to compile the document, but we ask you to:
\bibliographystyle{aa}     % style aa.bst
\bibliography{ms_tio_ia} % your references Yourfile.bib

\end{document}